\documentclass[letterpaper,twocolumn,prl,superscriptaddress,aps,showpacs,10pt,floatfix]{revtex4-1}

\usepackage[pdftex]{graphicx} 
\usepackage{epstopdf}
\usepackage{verbatim}
\usepackage{amsmath}
\usepackage{color}
\usepackage{subfigure}
\usepackage{amsbsy}
\usepackage{wasysym}

\begin{document}

\title{Rare-earth triangular lattice spin liquid: 
a single-crystal study of YbMgGaO$_{4}$}

\author{Yuesheng Li}
\affiliation{Department of Physics, 
Renmin University of China, 
Beijing 100872, P. R. China}

\author{Gang Chen}
\email{chggst@gmail.com, 
former affiliation: Department of Physics, University of Toronto, Ontario, Canada, M5S1A7}
\affiliation{State Key Laboratory of Surface Physics, 
Center for Field Theory and Particle Physics, 
Department of Physics, Fudan University, 
Shanghai 200433, China}
\affiliation{Collaborative Innovation Center of Advanced Microstructures,
Fudan University, Shanghai, 200433, China}

\author{Wei Tong}
\affiliation{High Magnetic Field Laboratory, Hefei Institutes of Physical Science, Chinese Academy of Sciences, Hefei 230031,P. R. China}

\author{Li Pi}
\affiliation{High Magnetic Field Laboratory, Hefei Institutes of Physical Science, Chinese Academy of Sciences, Hefei 230031,P. R. China}

\author{Juanjuan Liu}
\affiliation{Department of Physics, 
Renmin University of China, Beijing 100872, P. R. China}

\author{Zhaorong Yang}
\affiliation{Key Laboratory of Materials Physics, 
Institute of Solid State Physics, 
Chinese Academy of Sciences, 
Hefei 230031, P. R. China}

\author{Xiaoqun Wang}
\affiliation{Department of Physics, 
Renmin University of China, Beijing 100872, P. R. China}
\affiliation{Department of Physics and astronomy, 
Innovative Center for Advanced Microstructures, 
Shanghai Jiao Tong University, Shanghai 200240, P. R. China}

\author{Qingming Zhang}
\email{qmzhang@ruc.edu.cn}
\affiliation{Department of Physics, 
Renmin University of China, Beijing 100872, P. R. China}
\affiliation{Department of Physics and astronomy, 
Innovative Center for Advanced Microstructures, 
Shanghai Jiao Tong University, Shanghai 200240, P. R. China}
\date{\today}

\begin{abstract}

YbMgGaO$_{4}$, a structurally perfect two-dimensional triangular lattice with
odd number of electrons per unit cell and spin-orbit entangled effective spin-1/2 
local moments of Yb$^{3+}$ ions, 
is likely to experimentally realize the quantum spin liquid ground state. 
We report the first experimental characterization of single crystal
YbMgGaO$_{4}$ samples.
Due to the spin-orbit entanglement, the interaction 
between the neighboring Yb$^{3+}$ moments  
depends on the bond orientations and is highly anisotropic in the spin space. 
We carry out the thermodynamic and the electron spin resonance measurements to 
confirm the anisotropic nature of the spin interaction as well as 
to quantitatively determine the couplings.
Our result is a first step towards the theoretical understanding of the  
possible quantum spin liquid ground state in this system 
and sheds new lights on the search of quantum spin liquids 
in strong spin-orbit coupled insulators.   

\end{abstract}


\maketitle

\emph{Introduction.}---Recent theoretical advance has extended the 
Hastings-Oshikawa-Lieb-Schultz-Mattis theorem  
to the spin-orbit coupled insulators~\cite{arXiv1505.04193,Hastings2004,Oshikawa2000,LSM}. 
It is shown that as long as the time reversal symmetry is preserved, the ground state 
of a spin-orbit coupled insulator with odd number of electrons per unit 
cell must be exotic~\cite{arXiv1505.04193}. This important result 
indicates that the ground state of strong spin-orbit coupled insulators 
can be a quantum spin liquid (QSL). QSLs, as we use here, are new 
phases of matter that are characterized by properties such as quantum number 
fractionalization, intrinsic topological order, and gapless excitations 
without symmetry breaking~\cite{Balents2010,Wen2007}. 
Among the existing QSL candidate materials~\cite{organics1,organics2,Morita2002,kappaET,organictherm,
dmit,cscucl,cscucl2,nigas,QSI1,
hyperk,voborthite,volborthite2,vesignieite,Kapellasite2012,
kagome,kagome_NMR,mendels2007,
Cheng2011,Han2012,TbTiO,Ross2011,li14,li2013structure,li2013transition,Bieri2015,CeSnO2015}, 
the majority have a relatively weak spin-orbit coupling (SOC), 
which only slightly modifies the usual
SU(2) invariant Heisenberg interaction by introducing weak 
anisotropic spin interactions such as Dzyaloshinskii-Moriya 
interaction~\cite{Moriya1960a,Moriya1960b,Dzialoshinski1958}. 
It is likely that the QSL physics in many of these systems 
mainly originates from the Heisenberg part of the Hamiltonian 
rather than from the anisotropic interactions due to the weak SOC.
The exceptions are the hyperkagome Na$_4$Ir$_3$O$_8$  
and the pyrochlore quantum spin ice materials where the 
non-Heisenberg spin interaction due to 
the strong SOC plays a crucial role in determining the ground state 
properties~\cite{hyperk,QSI1,Chen2008,Curnoe2008,YbTiO2,Onoda2010,YbTiO2012,
PrSnO,YbTiO2012b,Chen2013,PrZrO2013,Alicea2014,TbTiO3,Huang2014}, 
though both systems contain even number of electrons per unit cell. 
Therefore, it is desirable to have a QSL candidate system in the
spin-orbit coupled insulator that contains  
odd number electrons per unit cell, where the strong SOC 
leads to a non-Heisenberg spin Hamiltonian~\cite{Curnoe2008,Chen2008,WCKB,Jackeli2009,Onoda2010,Chen2010,Chen2011,Huang2014}. 

\begin{figure}[t]
\begin{center}
\includegraphics[width=6.5cm,angle=0]{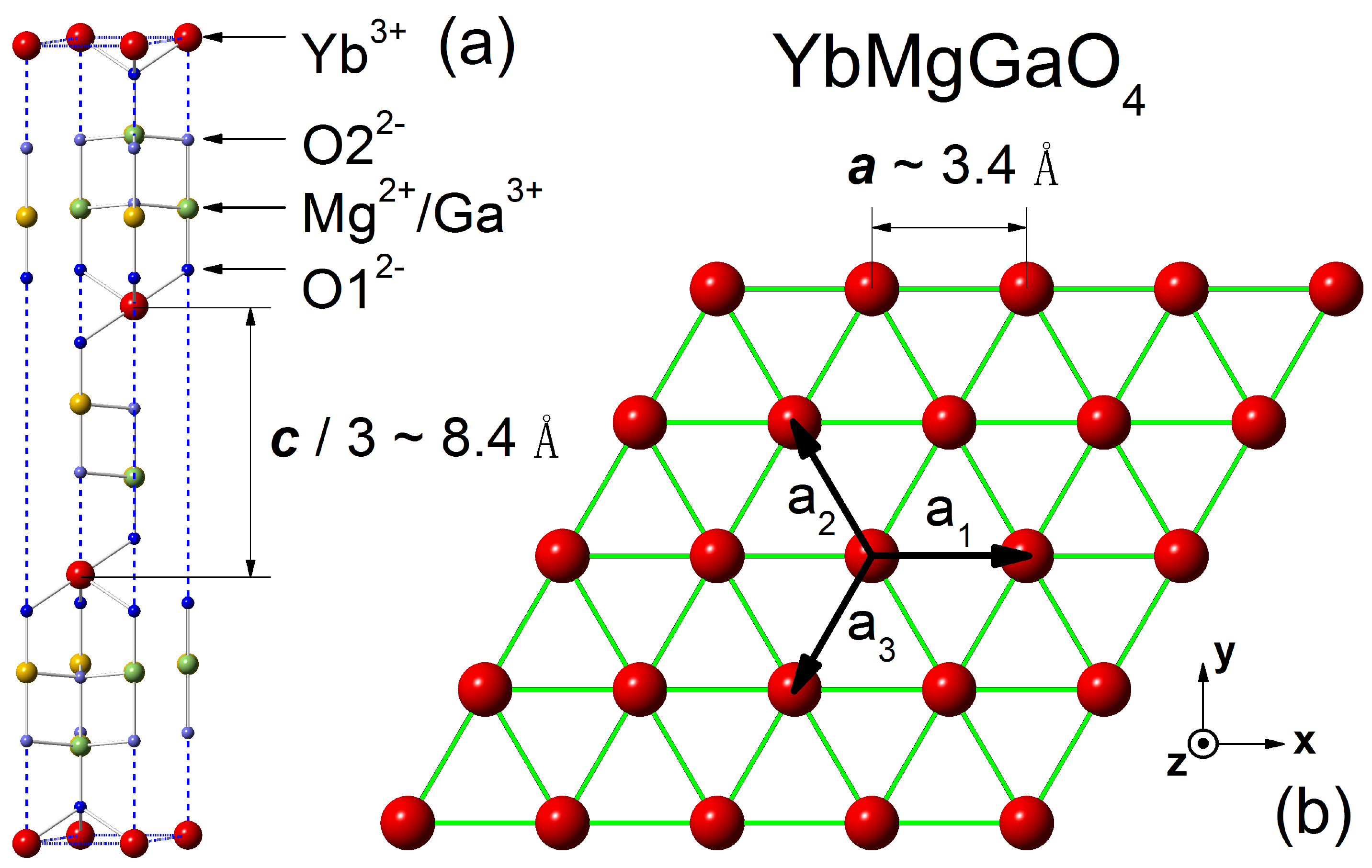}
\caption{(Color online.) The YbMgGaO$_{4}$ lattice structure (a) 
and the triangular lattice in the $ab$ plane (b) formed by the Yb$^{3+}$ 
ions. The inset defines the coordinate system for the spin components. }
\label{fig1}
\end{center}
\end{figure}

In this Letter, we propose a possible experimental realization of the  
QSL with strong SOC and odd number of electrons per unit cell 
in YbMgGaO$_{4}$, where the Yb$^{3+}$ ions form a perfect triangular 
lattice (see Fig.~\ref{fig1}). 
It was previously found in a powder sample that the system 
has a Curie-Weiss temperature 
$\Theta_{\text{CW}}^{\text{Powder}}\simeq -4$K
but does not order magnetically down to $60$mK~\cite{powderpaper}.  
To understand the nature of the obviously disordered ground state 
observed in YbMgGaO$_{4}$, it is necessary to have a 
quantitative understanding of the local moments and microscopic Hamiltoninan.
We here confirm the effective spin-1/2 nature of the Yb$^{3+}$ local moments
at low temperatures from the heat capacity and the magnetic entropy measurements
in high-quality single crystal samples. Because the Yb$^{3+}$ ion contains 
odd number of electrons, the effective spin is described by a Kramers' doublet. 
Based on this fact, we theoretically derive the symmetry allowed
spin Hamiltonian that is non-Heisenberg-like and involves four 
distinct spin interaction terms because of the strong SOC. 
Combining the spin susceptibility results along different crystallographic
directions and the electron spin resonance (ESR) measurements
in single crystal samples, we quantitatively confirm the anisotropic 
form of the spin interaction. We argue that 
the QSL physics in YbMgGaO$_{4}$ may originate from the 
anisotropic spin interaction. 
To our knowledge, YbMgGaO$_4$ is probably the first 
strong spin-orbit coupled QSL candidate system that 
contains odd number of electrons per unit cell with effective spin-1/2 local moments.

\emph{Experimental technique.}---High-quality single crystals ($\sim$ 1cm)
of YbMgGaO$_4$, as well as the non-magnetic iso-structural material
LuMgGaO$_4$~\cite{Supple}, are synthesized by the floating zone technique.
X-ray diffractions (XRD) are performed on the cutting single crystals to confirm the crystallization, the crystallographic orientation and the absence of the impurity phase, and for the single crystal structure refinements~\cite{gsas}.
The high quality of the crystallization was
confirmed by the narrow XRD rocking curves with $\Delta$2$\theta\sim 0.06^o$
and $0.04^o$ on $ab$ planes for YbMgGaO$_4$ and LuMgGaO$_4$ crystals, respectively.
Magnetization ($\sim$ 60mg of
YbMgGaO$_4$ single crystals) and heat capacity measurements (10 $\sim$ 20mg
of YbMgGaO$_4$ and LuMgGaO$_4$  single crystals) were performed using a
Quantum design physical property measurement system along and
perpendicular to the $c$ axis at 1.8 $\sim$ 400K under 0 $\sim$ 14T. The magnetic susceptibilities of single crystals agree with that of powder samples, $\chi_{\parallel}$/3+2$\chi_{\perp}$/3 $\simeq$ $\chi_{\text{Powder}}$.
The ESR measurements ($\sim$ 60mg of YbMgGaO$_4$
single crystals) at 1.8 $\sim$ 50K along different crystallographic
orientations were performed using a Bruker EMX plus 10/12 CW-spectrometer
at X-band frequencies (f $\sim$ 9.39GHz); the spectrometer was equipped
with a continuous He gas-flow cryostat.

\emph{Kramers' doublet and exchange Hamiltonian.}---The Yb$^{3+}$ ion 
in YbMgGaO$_4$ has an electron configuration $4f^{13}$, 
and from the Hund's rules
the orbital angular momentum ($L=3$) and the spin ($s=1/2$)
are entangled, leading to a total angular momentum $J=7/2$. Under the 
trigonal crystal electric field, the eight-fold degenerate 
$J=7/2$ states are splitted into four Kramers' doublets~\cite{Onoda2010,YbTiO2,Curnoe2008,Huang2014,YbTiO2012}.
By fitting the heat capacity results with an activated behavior, 
we find the local ground state doublet is well separated from the first 
excited doublet by an energy gap $\Delta \sim 420$K. 
This indicates that only the local ground state 
doublet is active at $T \ll \Delta$. Moreover, 
the magnetic entropy reaches to a plateau 
at R$\ln 2$ per mol Yb$^{3+}$ around 
$40$K, which is consistent with the 
thermalization of
the 2-fold degenerate ground state 
doublet~\cite{powderpaper,Supple}. 

As it is analogous to the local moments in the pyrochlore ice 
systems~\cite{TbTiO}, one can introduce an effective 
spin-1/2 degree of freedom, 
${\bf S}_i$, that acts on the local ground state doublet. The low-temperature 
magnetic properties are fully captured by these effective spins. 
Because the $4f$ electron is very localized spatially~\cite{Ross2011}, 
it is sufficient to keep only the nearest-neighbor interactions 
in the spin Hamiltonian~\footnote{The further neighbor magnetic dipole
interaction is estimated to be much smaller 
than the nearest neighbor couplings.}. 
Via a standard symmetry analysis, 
we find the generic spin Hamiltonian that is invariant under 
the R$\bar{3}$m space group symmetry of YbMgGaO$_4$ is given by 
\begin{eqnarray}
{\mathcal H} &=& \sum_{\langle ij \rangle}
\big[ 
J_{zz} S_i^z S_j^z + J_{\pm} (S^+_i S^-_j + S^-_i S^+_j)
\nonumber \\
&& + J_{\pm\pm} (\gamma_{ij} S^+_i S^+_j + \gamma_{ij}^{\ast} S^-_i S^-_j) 
\nonumber \\
&& - \frac{i J_{z\pm}}{2} (\gamma_{ij}^{\ast} S^+_i S^z_j 
- \gamma_{ij} S^-_i S^z_j
+ \langle i \leftrightarrow j \rangle )
\big],
\label{eqHam}
\end{eqnarray}
where $S^{\pm}_i = S^x_i \pm i S^y_i$, and
the phase factor $\gamma_{ij} = 1, e^{i 2\pi/3}, e^{-i2\pi/3}$
for the bond $ij$ along the ${\bf a}_1,{\bf a}_2,{\bf a}_3$ 
direction (see Fig.~\ref{fig1}), respectively. 
This generic Hamiltonian includes all possible microscopic 
processes that contribute to the nearest-neighbor spin interaction. 
The highly anisotropic spin interaction in ${\mathcal H}$
is a direct consequence of the spin-orbit entanglement 
in the local ground state doublet. Moreover, 
the antisymmetric Dzyaloshinskii-Moriya
interaction is prohibited in the Hamiltonian 
because of the inversion symmetry. 

\begin{figure}[t]
\begin{center}
\includegraphics[width=8.5cm,angle=0]{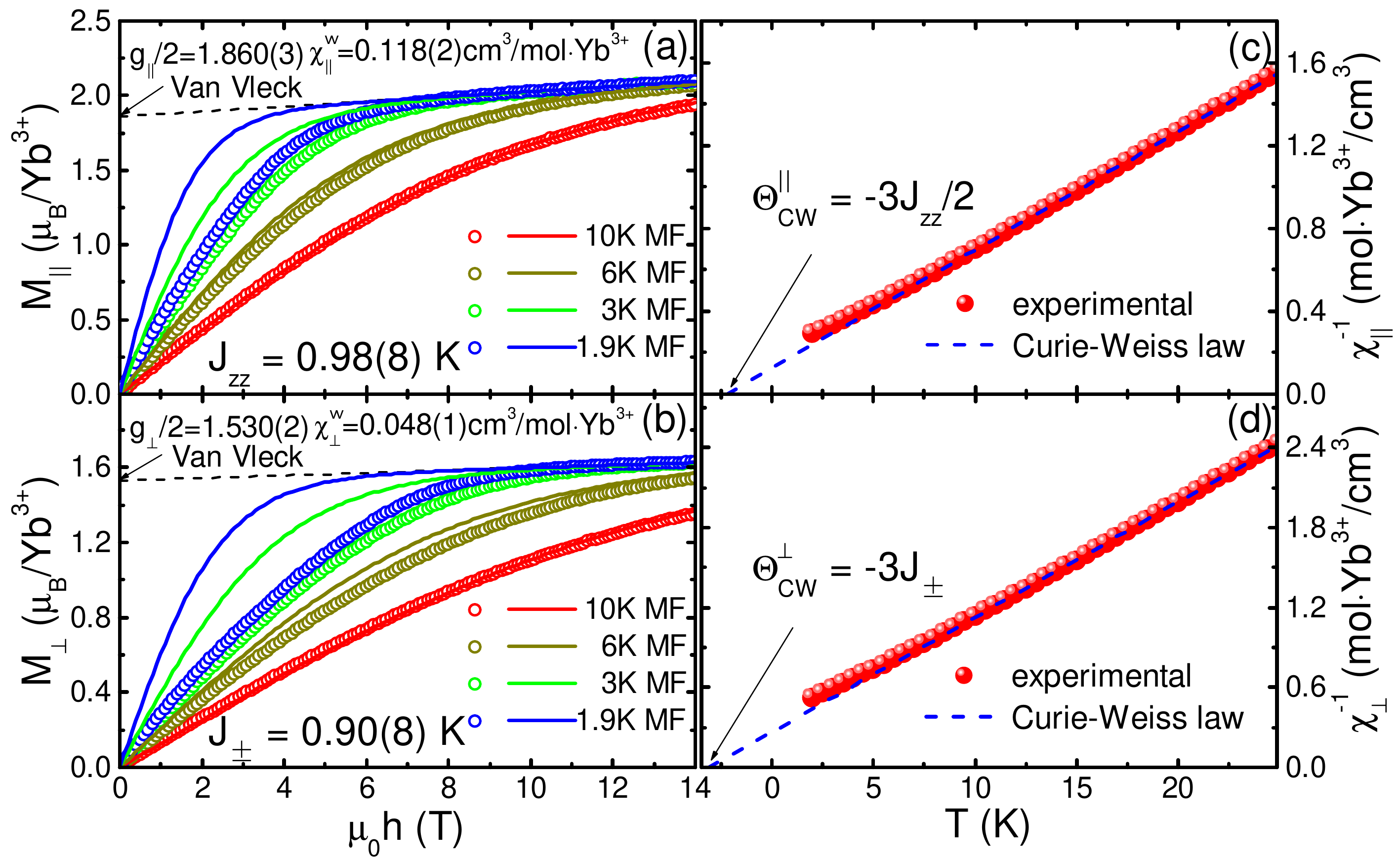}
\caption{(Color online.)
(a, b) The magnetization of the YbMgGaO$_4$
single crystals measured at 10, 6, 3 and 1.9K.
The dashed lines are linear fits of the
experimental results for fields above 12T at 1.9K.
The solid curves are the corresponding magnetization
calculated by the molecular field approximation. 
(c, d) The inverse spin susceptibilites (after 
subtracting the Van Vleck paramagnetism) fitted 
by the Curie-Weiss law (in dashed lines) 
for the YbMgGaO$_4$ single crystals.}
\label{fig2}
\end{center}
\end{figure}

\emph{Magnetization and magnetic susceptibility.}---In order to 
quantitatively determine the exchange couplings, we first 
perform the magnetization measurements for the YbMgGaO$_4$ single 
crystals down to $1.8$K under an external magnetic field 
(from $0$T to $14$T) parallel and perpendicular to 
the $c$ axis (see Fig.~\ref{fig1}). 
The Zeeman coupling to the external field 
is also constrained by the lattice symmetry and 
is given by~\footnote{$h_{\perp} \equiv (h_x^2 + h_y^2)^{1/2}$}
\begin{equation}
{\mathcal H}_{\text{Z}} = -\mu_0 \mu_{B}  \sum_i 
\big[ g_{\perp}  (h_x S^x_i +h_y S^y_i) 
+ g_{\parallel}  h_{\parallel} S^z_i
 \big].
 \label{Zeeman}
\end{equation}
As shown in Fig.~\ref{fig2}, the magnetization processes are 
obtained for both field directions at $1.9$K. When the field is above $12$T,
both magnetizations saturate and become linearly dependent on the field.
The slope of the $M$-$H$ curve is temperature-independent and is 
understood as the Van Vleck susceptibility 
($\chi_{\parallel}^{\text{VV}} = 0.118(2)$cm$^3$/mol$\cdot$Yb$^{3+}$,
$\chi_{\perp}^{\text{VV}} = 0.0479(8)$cm$^3$/mol$\cdot$Yb$^{3+}$) 
that arises from
the field-induced electronic transitions~\cite{shirata12}. 
After subtracting the Van Vleck paramagnetic contribution, 
we obtain the saturated magnetic moments 
($g_{\parallel}\mu_{\text B}/2$
and $g_{\perp}\mu_{\text B}/2$), 
from which we extract the $g$ factors $g_{\parallel} = 3.721(6), 
g_{\perp} = 3.060(4)$~\cite{Supple}.

We apply a small external field (0.01T) to measure the spin susceptibilities
parallel and perpendicular to the $c$ axis as a function of temperature. 
At high temperatures ($T \gtrsim 8$K) both susceptibilities
(after the substraction of Van Vleck paramagnetism) are well fitted 
by the Curie-Weiss law (see Fig.~\ref{fig2}). 
From the spin Hamiltonian, it is ready to obtain the 
Curie-Weiss temperatures $\Theta_{\text{CW}}^{\parallel} = - 3J_{zz}/2$  
($\Theta_{\text{CW}}^{\perp} = -3J_{\pm}$)      
for the field parallel (perpendicular) to the $c$ axis. 
We then use the above relations to find $J_{zz}$ and $J_{\pm}$. 
Alternatively, we apply the high-temperature molecular field approximation to 
fit the field dependence of magnetizations. 
As shown in Fig.~\ref{fig2}, the molecular field 
result agrees with the experiments
very well at 10K. These two approaches together 
yield $J_{zz}= 0.98(8)$K and $J_{\pm} = 0.90(8)$K. 
 
\begin{figure}[t]
\begin{center}
\includegraphics[width=8.5cm,angle=0]{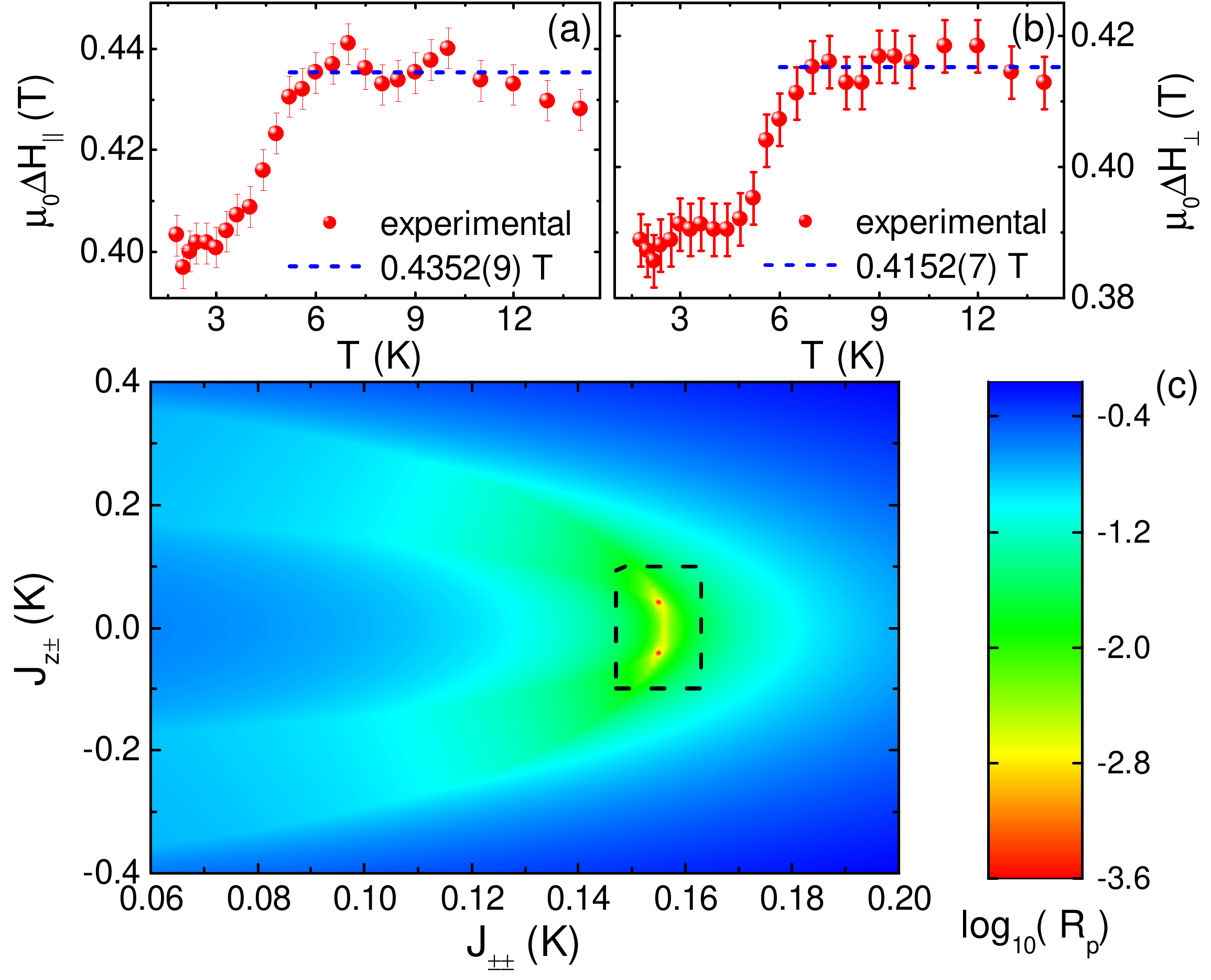}
\caption{(Color online.) 
The temperature dependence of ESR linewidths (a) parallel 
and (b) perpendicular to the $c$ axis. The dashed lines 
are the corresponding constant fits to the ESR linewidth 
data at  $T>6$K. (c)  The deviation, R$_p$, of the 
experimental ESR linewidthes from the theoretical ones 
for YbMgGaO$_4$. The dashed rectangle gives the optimal 
parameters $|J_{\pm\pm}| = 0.155(9)$K and $|J_{z\pm}| = 0.04(10)$K.}
\label{fig3}
\end{center}
\end{figure}

\begin{figure}[t]
\begin{center}
\includegraphics[width=8.cm,angle=0]{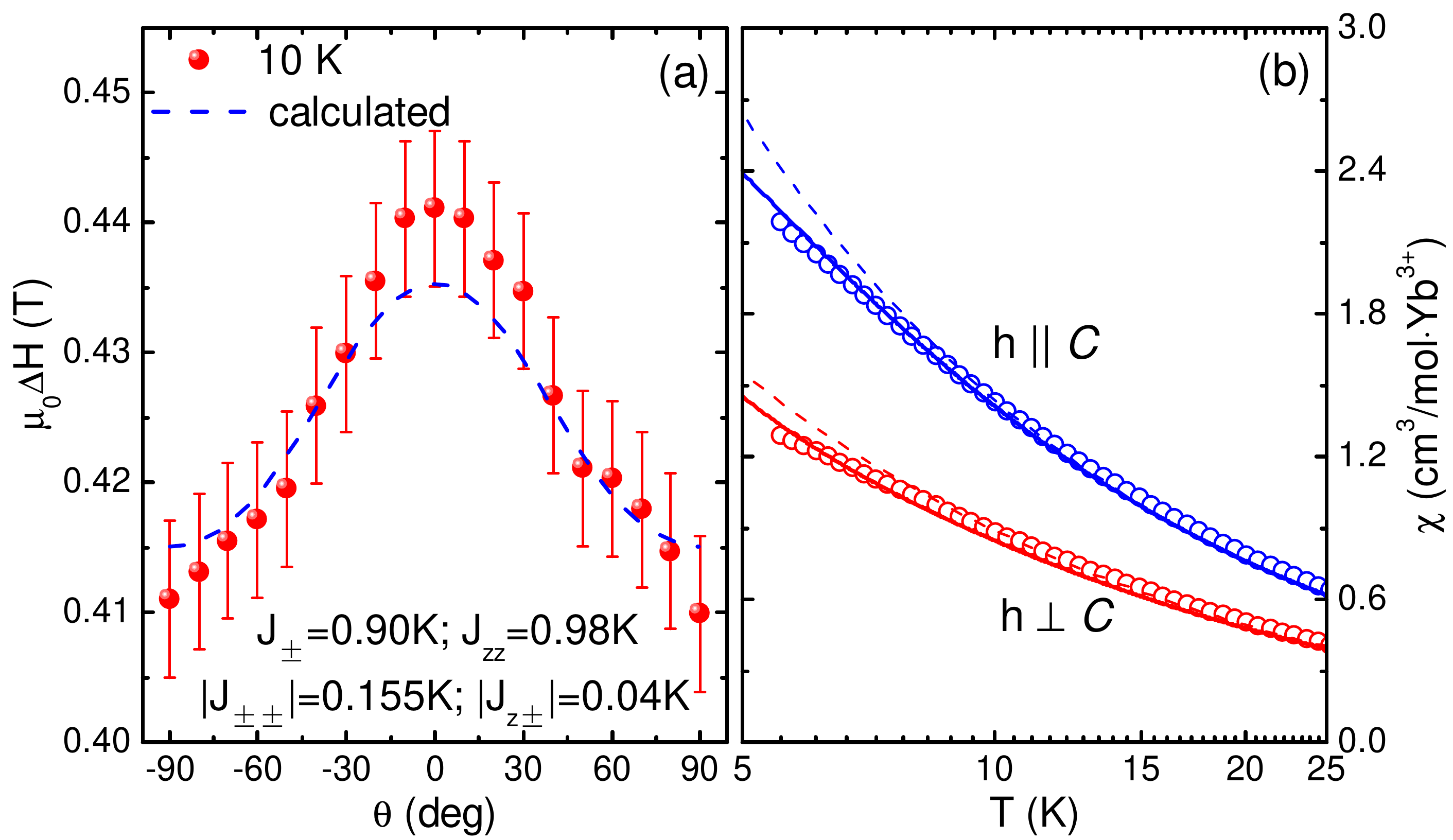}
\caption{(Color online.) (a) Angular dependence of ESR linewidth. 
The dashed curve is the calculated ESR linewidth. (b) The magnetic 
susceptibilities of YbMgGaO$_4$ single crystal after subtracting the
 Van Vleck paramagnetism. The solid curves are the calculated 
 susceptibilities using the high-temperature series expansion. 
 The dashed curves are the Curie-Weiss susceptibilities.}
\label{fig4}
\end{center}
\end{figure}

\emph{ESR.}---The remaining two coupling constants, $J_{\pm\pm}$ and $J_{z\pm}$,
that contribute to the anisotropic spin interaction,
completely break the U(1) spin rotation but keep time reversal symmetry
intact. They cannot be well resolved by the above thermodynamic measurements.
To precisely determine them, we apply the exhaustive ESR
measurements and analyze the ESR linewidths.
It is well-known that the ESR linewidth is a powerful
and direct measure of the anisotropic spin interactions~\cite{anderson53,castner71,Soos77,zorko08,pilawa97,nidda02}.
We perform the ESR measurements from 1.8K to 50K along
different crystallographic orientations, where the wide ESR signals,
as broad as $\mu_0 \Delta H(\theta) \sim 0.4$T, were observed
(see Fig.~\ref{fig3},~\ref{fig4} and raw ESR signals~\cite{Supple}).

Here we discuss various sources that broaden the ESR 
linewidth. The first one is the hyperfine interactions 
that contribute to the ESR linewidth with 
$\mu_0\Delta H_h \sim |A_{\parallel} |^2/(g \mu_B|J_0|) 
\sim 2$mT~\cite{pilawa97}, 
where the hyperfine coupling, $|A_{\parallel}|$, 
is about 2GHz for Yb$^{3+}$~\cite{misra98}, 
and $J_0$ is the isotropic Heisenberg coupling defined as
$J_0 \equiv (4J_{\pm}+J_{zz})/3 \sim 1.5(1)$K in Eq.~\eqref{eq7}. 
The next-nearest-neighbor magnetic dipole-dipole interactions also broaden the ESR signal
with $\mu_0 \Delta H_d \sim |E_{d}|^2/(g\mu_B|J_0|) 
\sim 0.3$mT~\cite{pilawa97}. 
Here, we have made a maximal estimate of the next-nearest-neighbor dipole-dipole 
interaction $|E_{d}|$ as $\mu_0 g^2 \mu_B^2/[4\pi (\sqrt{3} a)^3] $, 
where $a$ is the lattice constant. All the Yb$^{3+}$ ions 
share the same $g$-tensor, the Zeeman interaction is homogeneous 
and thus does not contribute to the ESR linewidth~\cite{pilawa97}. 
All the above contributions together give an
ESR linewidth that is two orders of magnitude smaller than the 
observed value. To account for such a large ESR linewidth 
that is $\sim 0.4$T, the only remaining origin lies in 
the anisotropy of the nearest-neighbor spin interaction. 

We now decompose the spin Hamiltonian 
in Eq.~\eqref{eqHam} into the isostropic and the anisotropic parts 
\begin{equation}
{\mathcal H} =J_{0}\sum_{\langle i,j \rangle} 
{\bf S}_i\cdot {\bf S}_j +{\mathcal H}' 
\label{eq7}
\end{equation}
where $J_0$ was previously introduced, 
$\Gamma_{ij}$ is a traceless coupling matrix,
and $\mathcal{H}'$ = $\sum_{\langle i,j \rangle}
{S}_i^{\mu} \Gamma_{ij,\mu\nu} {S}_j^{\nu}$
is the anisotropic part of the spin interaction.  
With the Zeeman term in Eq.~\eqref{Zeeman},
the ESR linewidth is obtained as 
\begin{equation}
\label{eq8}
\Delta H(\theta)=\frac{({2\pi})^{\frac{1}{2}}}{\mu_B g(\theta)}
\big({\frac{M_2^3}{M_4}}\big)^{\frac{1}{2}}
\end{equation}
where $\theta$ is the angle between the external field and the $c$ axis, 
$g(\theta) = ({g_{\parallel}^2 \cos^2\theta+g_{\perp}^2 \sin^2\theta})^{1/2}$, 
 $M_2 = \langle [{\mathcal H}',M^+][M^-,{\mathcal H}']\rangle/
 \langle M^+ M^-\rangle$ is the second moment, 
and $M_4 = \langle [{\mathcal H},
[{\mathcal H}',M^+]][{\mathcal H},[{\mathcal H}',M^-]]
\rangle/\langle M^+M^-\rangle$ is the fourth moment~\cite{zorko08}. 
Here, $M^{\pm} \equiv \sum_i S_i^{\pm}$.

The ESR signal of YbMgGaO$_4$ single crystal can be well fitted 
by the first-derivative Lorentzian line shape with a small 
contribution of dispersion as described by Ref.~\onlinecite{nidda02}. 
Both $\mu_0 \Delta H_{\parallel} (T)$ and 
$\mu_0 \Delta H_{\perp}(T)$ show a gradual broadening~\cite{Soos77} 
with increasing temperature for $k_B T < 5J_{0}$, and reach 
almost temperature-independent maxima  
at $\mu_0 \Delta H_{\parallel} = 0.4352(9)$T 
and $\mu_0 \Delta H_{\perp} = 0.4152(7)$T
for $k_B T \geq 5J_{0}$ (see Fig.~\ref{fig3}). 
We fit these high-temperature ESR linewidths 
according to the theoretical results.  
In Fig.~\ref{fig3}, we plot the deviation of the experimental result
from the theoretical one,
\begin{equation}
R_p = \frac{1}{2}\Big[ \Big|\frac{\Delta H_{\parallel}
-\Delta H_{\parallel}^{cal}}{\Delta H_{\parallel}}\Big|
+ \Big|\frac{\Delta H_{\perp}-\Delta H_{\perp}^{cal}}
{\Delta H_{\perp}} \Big|\Big],
\end{equation}
as a function of $J_{\pm\pm}$ and $J_{z\pm}$. 
The optimal fitting is obtained by setting
$|J_{\pm\pm}| = 0.155(3)$K and $|J_{z\pm}| = 0.04(8)$K
whose signs cannot be fixed by the fitting.
 
As an unbiased check of the fitted results, we use the 
optimal couplings to calculate the angle 
dependence of the ESR linewidth, 
$\mu_0 \Delta H^{cal}(\theta)$, where $\theta$ 
is the angle between the external field and
the $c$ axis. As shown in Fig.~\ref{fig4}, the experimental 
curve agrees with the theoretical result very well. 
Moreover, we apply the high-temperature series expansions 
to compute the spin susceptibilities per Yb$^{3+}$ ion up 
to $\mathcal{O}(T^{-3})$, 
\begin{eqnarray}\label{eq9}
\chi_{\parallel}&=&\frac{\mu_0 g_{\parallel}^2\mu_B^2}{4k_BT}(1-
\frac{3J_{zz}}{2k_BT}
 -\frac{3J_{\pm}^2 + J_{\pm\pm}^2 + J_{z\pm}^2}{2k_B^2T^2}
 \nonumber \\
&&  +   \frac{15 J_{zz}^2}{8 k_B^2 T^2}), 
 \\
\chi_{\perp}&=&\frac{\mu_0 g_{\perp}^2\mu_B^2}{4k_BT}(1-\frac{3J_{\pm}}{k_BT}
+\frac{7J_{\pm}^2}{k_B^2T^2} -\frac{2J_{\pm\pm}^2}{k_B^2T^2}
\nonumber 
\\
&& -\frac{5J_{z\pm}^2}{16k_B^2T^2}-\frac{J_{zz}^2}{8k_B^2T^2}
-\frac{J_{\pm}J_{zz}}{4k_B^2T^2}).
\end{eqnarray}
As we depict in Fig.~\ref{fig4}, the high-temperature expansion 
shows an better fitting with the experimental results 
at lower temperatures than the simple Curie-Weiss laws. 

\emph{Discussion.}---In the previous powder sample measurements, 
the magnetic heat capacity of YbMgGaO$_4$ behaves as 
$C_v \propto T^{\gamma}$ ($\gamma \approx 2/3$) from about $1$K 
down to 0.06K~\footnote{We have also measured the 
heat capacity for the single crystal down to 
0.06K, and it does not show any significant difference 
from the powder sample results.}, 
suggesting the system is probably
in a gapless QSL phase~\cite{powderpaper}.
The residual magnetic entropy of the system at 0.06K is 
less than 0.6$\%$ of the total magnetic entropy~\cite{powderpaper}. 
This is a strong indication that we are indeed accessing the ground state 
property. 
As far as we are aware of, this is the first clear observation 
of $C_v \propto T^{2/3}$ in QSL candidate systems. 
In fact, this behavior is compatible with what one may expect 
for the U(1) QSL with a spinon Fermi surface 
in two dimensions~\cite{Motrunich2005,Lee05,Motrunich2006},
a state previously proposed for the organic 
$\kappa$-(ET)$_2$Cu$_2$(CN)$_3$ and EtMe$_3$Sb[Pd(dmit)$_2$]$_2$~\cite{Motrunich2005,Lee05,Motrunich2006}.
Although alternative proposals also exist~\cite{Mishmash2013}, 
the QSL physics in the organics is believed to 
originate from the strong charge fluctuation of the weak 
Mott regime that induces a sizable ring exchange and thus 
destabilizes the 120$^o$ magnetic order for a triangular system~\cite{Motrunich2005,Motrunich2006,Lee05}. 
In contrast, the physical mechanism to realize possible
QSL in YbMgGaO$_4$ should be rather different. 
The $f$ electrons of YbMgGaO$_4$ are very localized and 
are in the strong Mott regime. 
The charge fluctuation is very weak and 
the ring exchange process should be negligible. 
On the other hand, the anisotropic $J_{\pm\pm}$ 
and $J_{z\pm}$ spin interaction is 
a new ingredient brought by the spin-orbit 
entanglement of the Yb $f$ electrons
and is expected to be the physical origin of 
the QSL physics. This is because in the absence of the 
anisotropic $J_{\pm\pm}$ and $J_{z\pm}$ spin interaction
the antiferromagnetic XXZ model would 
produce a conventional magnetic order~\cite{XXZ2014}. 
It is the anisotropic $J_{\pm\pm}$ and $J_{z\pm}$ 
spin interaction that competes with the 
XXZ model and may melt the magnetic order in certain parameter regime~\cite{Chen2015}. 
Through the current single crystal measurements, 
we expect YbMgGaO$_4$ to be a spin-orbit coupled QSL in which the 
anisotropic spin interaction is the driving force.

To summarize, we have characterized the magnetic properties of
large YbMgGaO$_4$ single crystals that are grown for the first time.  
The crystal structure and effective spin-1/2 Hamiltonian of YbMgGaO$_4$ 
are precisely determined by single crystal X-ray diffractions, thermodynamic 
measurements and ESR linewidths on the orientated single crystals. 
We find that the anisotropic spin exchange interaction on the Yb triangular 
lattice significantly broadens the ESR linewidths.
We argue that the anisotropic spin interaction plays an important role 
to stabilize the possible QSL ground state in YbMgGaO$_4$. In the future, 
it will be interesting to numerically study the theoretical model in our work, 
classify QSLs in strong spin-orbit coupled insulators, and use 
inelastic neutron scattering to detect the possible fractionalized 
spin excitation in the single crystal samples.

\emph{Acknowledgement.}---We thank Rong Yu for helpful conversation. 
This work was supported by the NSF of China and 
the Ministry of Science and Technology of China 
(973 projects: 2011CBA00112 and 2012CB921701). 
G.C. was supported by the starting up funds of Fudan University.  
Q.M.Z. and Y.S.L. was supported by the Fundamental Research 
Funds for the Central Universities, and the Research Funds 
of Renmin University of China.

\bibliography{YMGO}

\end{document}